\documentclass[twocolumn]{aastex631}

\usepackage{bm}
\usepackage{mathtools}
\usepackage{graphicx}

\shorttitle{AI-Assisted Synthetic Astrophysicists}

\begin{document}

\title{AI Cosplaying as Astrophysicists: A Controlled Synthetic-Agent Study of AI-Assisted Astrophysical Research Workflows}

\author[0000-0001-6406-1003]{Chun Huang}
\affiliation{Department of Physics and McDonnell Center for the Space Sciences, Washington University in St. Louis, St. Louis, MO 63130, USA}
\correspondingauthor{Chun Huang}
\email{chun.h@wustl.edu}

\begin{abstract}
Large Language Models (LLMs) are now widely used in astrophysics, but do they actually make our lives easier, or do they merely invent new physics with enough confidence to hide a minus sign? In a specialized field where checking fluent hallucinations is itself labor-intensive, AI assistance can demand as much work as the task it claims to simplify. To evaluate where AI genuinely improves scientific workflows, we bypassed human trials and instead forced AI agents to cosplay as astrophysicists. We simulated 144 synthetic researchers, varying in career stage, AI awareness, and willingness to verify outputs, across 2,592 daily astrophysics research assignments. Comparing solo work against four styles of AI assistance produced 12,960 scored episodes. No assisted policy universally outperformed unassisted work in the primary Qwen production run. Instead, performance depends strongly on the task, the style of AI use, and the identity of the actor. While cautious assistance helps on creative, extractive, and critique-oriented tasks, it can fail catastrophically on derivation-heavy physics. A full actor-swap DeepSeek rerun changes that picture materially: verification-heavy use becomes the strongest assisted policy, two assisted modes enter the higher-utility/lower-risk quadrant, and the derivation-heavy fragility that dominates the Qwen production run largely disappears. In its current form, AI is useful, but only conditionally, its value is uneven, task-specific, and shaped jointly by workflow, usage policy, and which LLM you are using.
\end{abstract}

\section{Introduction}

We are living through a genuine AI transition. With advances in transformer-based models arriving on timescales that feel measured in months rather than years, large language models (LLMs) are already reshaping scientific practice \citep{Kusumegi_2025,zhang2025,luo2025llm4srsurveylargelanguage}. Whether we welcome this shift or not, it is unlikely to leave our workflows unchanged. Yet the response within the scientific community has been far from uniform. Alongside enthusiasm, many colleagues have expressed understandable caution about how these tools should be used responsibly and where their legitimate role in research truly begins and ends \citep{ma_2025,vanDis2023ChatGPTFP,Thorp_2023,hogg2026astrophysics}. That caution is well founded. Foundation models may assist at many stages of the research cycle, including literature synthesis, drafting, coding, data interpretation, and even hypothesis generation, while reliability, grounding, and evaluation remain unresolved core challenges \citep{WangEtAl2023,zhang2025}. Beyond science-specific settings, controlled and field studies have shown that generative-AI assistance can substantially improve throughput in professional writing, customer support, and some coding tasks, although these gains are heterogeneous, uneven, and sharply task-dependent rather than universal \citep{NoyZhang2023,BrynjolfssonEtAl2023,PengEtAl2023}. Scientific work, however, is less forgiving. Here, a fluent but incorrect answer does not merely weaken presentation, a missing minus sign can change the conclusion, a polished paragraph can hide a bad assumption, and a convincing explanation can occasionally invent new physics while failing basic algebra. For that reason, differences in performance (heterogeneity) are not secondary to the question of whether AI is useful in science. They are central to it. 

Astronomy and astrophysics are already participating in the broader AI transition. Recent community studies and astronomy-specific LLM projects show that researchers are exploring these systems for literature discovery, synthesis, hypothesis generation, coding, drafting, and outreach, while consistently stressing that outputs must be validated against domain knowledge and the underlying literature before they can be trusted \citep{Iyer_2024,ciuca2023galacticchitchatusinglarge,perkowski2024astrollamachatscalingastrollamaconversational,FouesneauEtAl2024,HykEtAl2025}. In parallel, astronomy-specific evaluation efforts have expanded rapidly. These include benchmark-driven studies of factual and conceptual knowledge, literature-oriented interaction studies, and workflow evaluations centered on code and visualization  \citep{WuEtAl2024,TingEtAl2025,JosephEtAl2025,ciuca2023galacticchitchatusinglarge,HykEtAl2025}. Recent population-level evidence also suggests a measurable rise in LLM-mediated writing in astronomy papers and science papers more broadly \citep{AstaritaEtAl2024,Liang2025QuantifyingLL}. Therefore, the question is no longer whether LLMs are relevant to astrophysical research. The question now is how they should be used, where they are genuinely helpful, and where they are still likely to fail in highly specialized astronomy workflows.

Taken together, the existing studies are informative, but they still leave an important gap. Most current evaluations examine only one part of the problem at a time, like benchmark-style tests of astronomy knowledge, literature-centered interactions, small studies of how astronomers actually use and judge these systems, or workflow evaluations focused on code, data analysis, and visualization. What is still missing is a unified view of how assistance performs across different kinds of researchers, tasks, and usage styles \citep{WuEtAl2024,TingEtAl2025,HykEtAl2025,JosephEtAl2025,FouesneauEtAl2024,Iyer_2024}. More general scientific benchmarks such as GPQA are useful for testing difficult expert-level reasoning, but they do not, by themselves, capture the mixed workflow reality of astrophysics research. In practice, astrophysicists move across a diverse set of tasks, including concise but technical writing, email drafting, literature search and extraction, coding and debugging, data analysis and visualization, critical evaluation of claims, and open-ended planning under uncertainty \citep{ReinEtAl2024,FouesneauEtAl2024,Iyer_2024,HykEtAl2025}. Running broad human-subject productivity trials is possible, but standardizing them is difficult. Most studies so far examine one domain at a time, which is valuable for causal identification \citep{NoyZhang2023,BrynjolfssonEtAl2023,PengEtAl2023,luo2025llm4srsurveylargelanguage,fragiadakis2025evaluatinghumanaicollaborationreview} but leaves open the larger workflow question --- how performance changes when the same assignments are completed with no assistance, with assistance, and with different ways of using that assistance. That is the gap a controlled workflow-level experiment can address.

In this paper, we address that gap with a controlled synthetic-agent experiment tailored to astrophysical research workflows. This is not an attempt to claim that model-based agents are equivalent to human astronomers. It should be viewed as a numerical protocol designed to estimate matched within-task contrasts under fixed and auditable conditions. Our design treats AI use not as a binary intervention, but as a policy-sensitive choice: cautious assistance, low-verification assistance, verification-heavy assistance, and overtrusting assistance are not the same AI usage in science practice. In the main experiment, a balanced population of 144 synthetic researcher profiles is evaluated on 2592 distinct tasks drawn from a broad astrophysics reservoir spanning various workflow families. Each assignment is executed once in a solo condition and once under each of four assisted usage styles, yielding 12960 scored episodes in the main production run with one open-source LLM. We pair this main experiment with a full actor-swap rerun under the same scoring framework so that cross-model robustness can be assessed on the same canonical design.

Section \ref{sec:methods} describes the numerical experiment design, scoring framework, and analysis pipeline. Section \ref{sec:results} presents the main experimental results, including population-level conclusions across the agent ensemble. Section \ref{sec:deepseek_robustness} discusses the effects of the model-swap rerun on the main conclusions. Section \ref{sec:conclusion} discusses the limitations of the current study and closes with our overall conclusions and directions for future work.

\section{Methods}
\label{sec:methods}

In this work, we mainly study AI-assisted astrophysics workflows as a controlled synthetic-agent numerical experiment rather than as a human-subject productivity trial. The objective is to measure how assistance policy changes the quality of completed scientific work, the frequency of severe failures, and the heterogeneity of these effects across researcher profiles and task families under a fixed and auditable design. The main analysis reported in the paper is the completed large Qwen-based run. In addition, we describe a DeepSeek cross-model validation run that changes the actor model while holding the task and scoring framework fixed across the full canonical design.

\begin{figure*}[t]
    \centering
    \IfFileExists{figure1_demo.pdf}{
        \includegraphics[width=\linewidth]{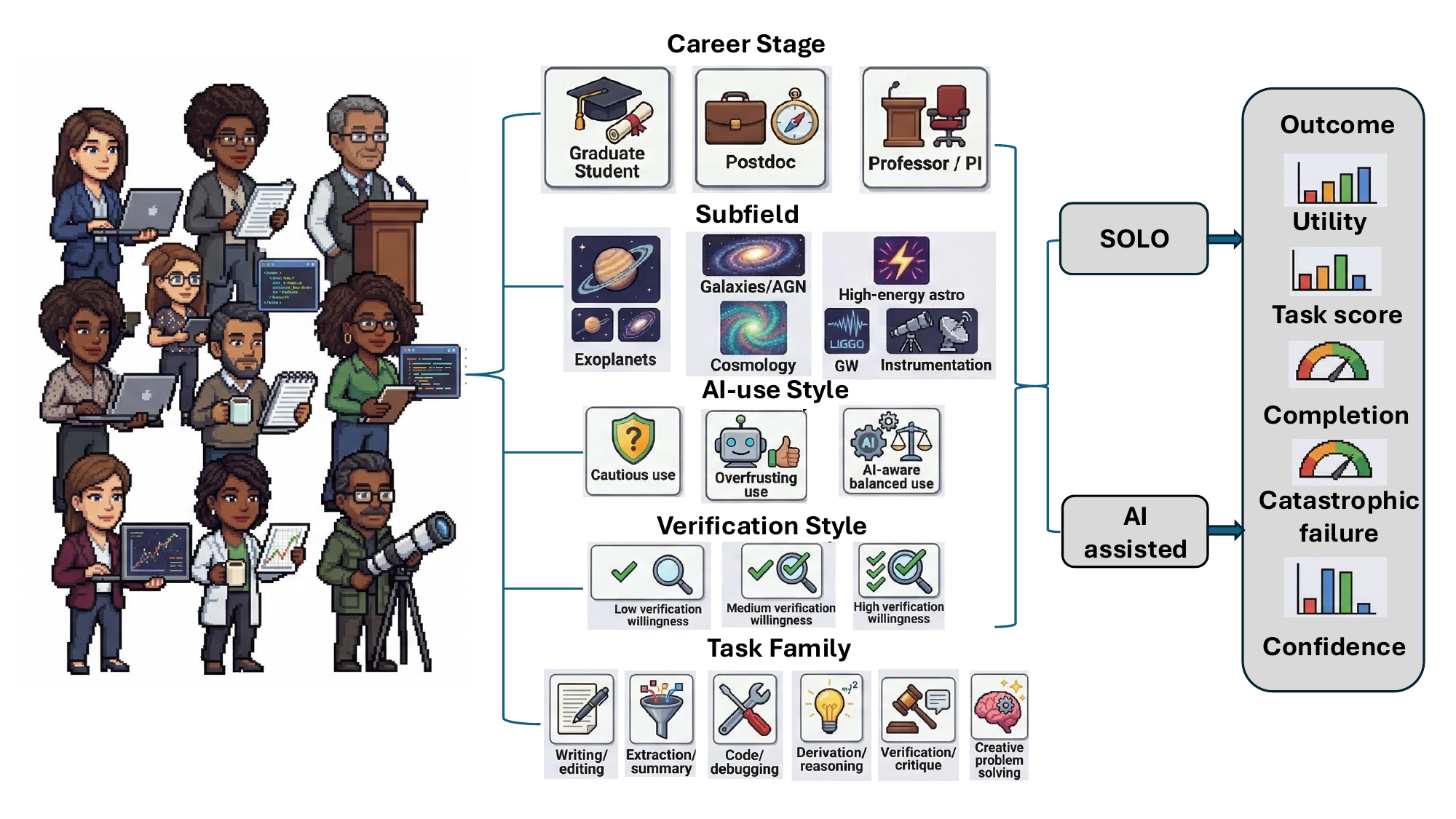}
    }{
        \fbox{\parbox[c][0.26\textheight][c]{0.92\linewidth}{\centering Figure 1 placeholder\\Upload \texttt{figure1\_demo.pdf} to Overleaf.}}
    }
    \caption{Overall workflow of the synthetic-agent experiment. A balanced population of AI agent astrophysicists is combined with a broad astrophysics task reservoir and a precomputed assignment table before any model calls are made. Each assignment is then executed under one \texttt{solo} condition and four assisted usage styles, scored under a common judging framework, and aggregated into matched assisted-versus-solo contrasts for the core outcome summary, heterogeneity analyses, usage-style frontier, and cross-model validation.}
    \label{fig:workflow}
\end{figure*}

Figure~\ref{fig:workflow} summarizes the numerical workflow at the highest level. The experiment begins with construction of a balanced synthetic population of astrophysicists and a task bank designed to cover multiple astrophysical subfields and workflow families. Before any model calls are made, these ingredients are combined into a reproducible assignment table that fixes which agent sees which task under which replicate. Each assignment is then executed under one solo condition and four AI-assisted usage styles. The resulting products are scored by a common judge, converted into episode-level metrics, and finally analyzed through matched assisted-versus-solo contrasts. The same workflow is reused for the DeepSeek cross-model rerun, with the actor model changed but the assignment and scoring framework held fixed.

\subsection{Production Large Synthetic-Agent Experiment}

The primary experiment used a synthetic population of $144$ agents. Each agent was defined by a career stage, an AI-awareness level, and a verification-willingness level. Career stage had four levels intended to capture different positions in the astrophysics research community: early graduate student, late graduate student, postdoctoral researcher, and faculty-like senior researcher. These were associated with monotone expertise scores on $[0,1]$,
\[
\begin{aligned}
\mathrm{early\_grad} &= 0.28, \qquad
\mathrm{late\_grad} = 0.46, \\
\mathrm{postdoc} &= 0.66, \qquad
\mathrm{faculty} = 0.82.
\end{aligned}
\]
Here AI awareness denotes the synthetic researcher's familiarity with the capabilities, limitations, and appropriate use cases of AI tools, rather than raw astrophysical expertise. In the real runs, this variable functions as a role-conditioning descriptor as well: higher AI awareness is intended to represent a researcher who is more able to use assistance strategically and to recognize when AI output deserves caution, but it does not mechanically modify the model's underlying capability. AI awareness was represented at three levels ($0.22$, $0.55$, $0.82$), and verification willingness at three levels ($0.18$, $0.52$, $0.85$). These numerical values were not interpreted as psychometric measurements, they were compact latent descriptors used to stabilize role prompting and to support structured heterogeneity analysis.

The $144$ agents were constructed on the complete profile grid,
\[
\begin{aligned}
4\ \mathrm{career\ stages}
&\times 3\ \mathrm{AI\!-\!awareness\ levels} \\
&\times 3\ \mathrm{verification\ willingness\ levels}
= 36\ \mathrm{cells},
\end{aligned}
\]
with exactly four agents in each cell. This design guarantees exact balance not only in the marginal distributions of career stage, AI awareness, and verification willingness, but also in the full joint profile structure. A representative synthetic profile might therefore be a \emph{late\_grad} agent with expertise $0.46$, \emph{high} AI awareness, and \emph{low} verification willingness, while another might be a \emph{faculty}-level agent with expertise $0.82$, \emph{medium} AI awareness, and \emph{high} verification willingness. These profiles define who the agent is supposed to be, while the assistance condition is imposed separately through the usage-style prompt.

The task reservoir was built to resemble ordinary astrophysics research work rather than benchmark-style puzzle solving. We defined six workflow families intended to capture common research activities: writing/editing, extraction/synthesis, code debugging, derivation/reasoning, creative problem solving, and verification/critique.

We then used a humanized broad-astrophysics bank of $3000$ self-contained tasks, with exactly $500$ tasks in each family and broad topical coverage across exoplanets and stellar astrophysics, galaxies and AGN, high-energy astrophysics, cosmology and large-scale structure, compact objects and gravitational-wave science, and instrumentation, survey, and time-domain astronomy. Each task carries structured metadata, including task family, astrophysical domain, difficulty, verifiability, ambiguity, and a coarse task-context label. In the final bank, task difficulty ranged from $0.22$ to $0.63$, verifiability from $0.68$ to $0.96$, and ambiguity from $0.12$ to $0.46$. Table~\ref{tab:workflow_families} gives some illustrative tasks from each family together with the practical logic used to rate them, with the six examples spanning distinct astrophysical subfields.

\begin{deluxetable*}{lll}
\tablecaption{ Illustrative tasks from the broad astrophysics reservoir and the corresponding family-specific rating logic. The examples are illustrative of the task style rather than verbatim prompts.\label{tab:workflow_families}}
\tablewidth{0pt}
\tablehead{
\colhead{Family} & \colhead{Illustrative harder task} & \colhead{How the task is rated}
}
\startdata
\parbox[t]{0.16\textwidth}{Writing/editing} &
\parbox[t]{0.28\textwidth}{Rewrite a target-of-opportunity request for rapid X-ray follow-up so that it is urgent, scientifically careful, and free of hype.} &
\parbox[t]{0.40\textwidth}{High score if the text motivates urgency, preserves the observed brightening and possible spectral hardening, and remains scientifically careful rather than inflated. Penalties apply if the wording overclaims or drops the scientific trigger for the request.} \\
\parbox[t]{0.16\textwidth}{Extraction/synthesis} &
\parbox[t]{0.28\textwidth}{Summarize what changed in a compact-object draft revision after rerunning the inference with a wider distance prior and adding a mass--radius corner plot.} &
\parbox[t]{0.40\textwidth}{High score if the summary captures the prior change, the new figure, and the softened physical interpretation, while staying concise and avoiding invented changes that were not in the revision note.} \\
\parbox[t]{0.16\textwidth}{Code debugging} &
\parbox[t]{0.28\textwidth}{Diagnose a catalog cross-match bug in which \texttt{sep\_arcsec < 2} was used even though the intended search radius was $2$ arcmin.} &
\parbox[t]{0.40\textwidth}{High score if the answer identifies the arcmin--arcsec mismatch and gives a valid correction, such as changing the threshold to $120$ arcsec or converting the separation units consistently. Severe penalties apply if the answer leaves the original unit bug unresolved.} \\
\parbox[t]{0.16\textwidth}{Derivation/reasoning} &
\parbox[t]{0.28\textwidth}{Estimate the Eddington ratio for a $10^8\,M_\odot$ black hole with $L_{\mathrm{bol}}=6.3\times10^{45}\,\mathrm{erg\,s^{-1}}$ using the standard $L_{\mathrm{Edd}}$ scaling and report the physically relevant ratio.} &
\parbox[t]{0.40\textwidth}{High score if the answer computes both $L_{\mathrm{Edd}}$ and $L_{\mathrm{bol}}/L_{\mathrm{Edd}}$ correctly, keeps track of the powers of ten, and states the final ratio clearly. Catastrophic failure is reserved for severe but fluent reasoning errors, for example ratios that are orders of magnitude wrong but presented with confidence.} \\
\parbox[t]{0.16\textwidth}{Creative problem solving} &
\parbox[t]{0.28\textwidth}{Advise a collaboration on how to decide whether to trigger broad follow-up for a compact-binary candidate with a large sky localization.} &
\parbox[t]{0.40\textwidth}{High score if the answer discusses realistic workflow considerations such as localization area, counterpart timescale, observability, and resource tradeoffs, rather than drifting into vague or grand-strategic language.} \\
\parbox[t]{0.16\textwidth}{Verification/critique} &
\parbox[t]{0.28\textwidth}{Critique the claim that a parameter shift after changing a cosmological scale cut is direct evidence for new physics beyond the baseline model.} &
\parbox[t]{0.40\textwidth}{High score if the answer raises analysis or systematic explanations before any appeal to new physics, and identifies the missing inferential step rather than merely repeating the claim in softer language.} \\
\enddata
\end{deluxetable*}

The production actor prompt was organized in two layers. A system message encoded the synthetic role, including career stage, expertise, AI-awareness level, verification-willingness level, experimental condition, and usage style. A user message then specified the task family, task context, difficulty, verifiability, ambiguity, and the task itself. The actor was required to return a final work product, a scalar confidence value, a short note describing verification steps actually performed, and a binary completion flag. This requirement that verification notes describe \emph{actual} checks rather than generic caution language was important to the design because it made the usage styles behaviorally observable in the output itself.

Each task was evaluated once in a \texttt{solo} condition and once under each of four assisted usage styles: \texttt{cautious\_assisted}, \texttt{low\_verification}, \texttt{verification\_heavy}, and \texttt{overtrusting}. These usage styles were designed to represent different ways of employing assistance rather than different classes of agents. In the current version, \texttt{cautious\_assisted} instructed the actor to use AI as a draft aid but to independently verify key claims and to follow the independent check rather than the draft if the two disagreed. \texttt{verification\_heavy} was still stricter, explicitly requiring re-derivation of equations, arithmetic checks, unit checks, and line-by-line code inspection when relevant. \texttt{low\_verification} allowed at most a lightweight sanity check and emphasized speed, whereas \texttt{overtrusting} instructed the actor to rely strongly on the draft and to intervene only when an error was obvious. These stronger policy prompts were adopted after a matched diagnostic study showed that earlier policy definitions were not behaviorally distinct enough in practice.

Operationally, these assisted conditions were implemented as explicit policy text inside the actor's system prompt, not as a literal two-stage architecture in which one model first produced a separate draft and the actor then revised it. In other words, the current pipeline makes a single actor-model call per episode and asks that actor to behave as if it were working with AI assistance under the specified policy. We chose this prompt-operationalized design deliberately. A literal helper-draft pipeline would introduce additional degrees of freedom, including helper-model identity, helper prompt, draft formatting, and how the draft was exposed to the actor, thereby conflating assistance policy with pipeline engineering. The present design instead keeps the interface, model stack, context structure, and latency fixed across conditions, makes matched policy contrasts cleaner, and remains computationally tractable at the large scale run. The current study should therefore be interpreted as evaluating policy-conditioned assistance behavior under a fixed actor interface rather than as evaluating a separate external-draft architecture.

Because these assisted policies are intentionally distinct, we do not average them into a single pooled assisted condition in the main-text analyses. Instead, Figures~\ref{fig:core_effects} and \ref{fig:heterogeneity} use \texttt{cautious\_assisted} as the primary assisted comparator. This policy is the most interpretable single-policy baseline for the main text because it represents a realistic checked-assistance workflow. In the full policy frontier reported in Figure~\ref{fig:frontier}, it is also the most favorable assisted compromise.

A central methodological choice was to precompute the full assignment table before any model calls were made. Each row in this table specified an assignment identity together with deterministic seeds, agent metadata, task metadata, task order, and audit fields describing whether any fallback logic had been used during assignment generation. The assignment algorithm enforced three constraints. First, tasks were globally unique within the run, so the main experiment used $2592$ distinct tasks from the $3000$-task reservoir. Second, task-family quotas were near-equal within each agent portfolio. Because the main run used $18$ tasks per agent and there were six workflow families, each agent received exactly three tasks from each family. Third, astrophysical domain coverage was balanced as strongly as possible at the study level by preferring underused domains within the available family-specific pools and only falling back to broader choices when a target cell was exhausted. All assignments were deterministic under a master seed of $7$.

The resulting precomputed table therefore contained
\[
144 \times 18 \times 1 = 2592
\]
assignment rows, and because each assignment was executed under five usage styles, the expected episode count was
\[
144 \times 18 \times 1 \times 5 = 12960.
\]
By construction, the assignment table was exactly balanced by task family, with $432$ assignments in each family. Domain counts were near-balanced but constrained by the raw task-bank composition, yielding $473$ assignments each for exoplanets/stellar, galaxies/AGN, high-energy astrophysics, and cosmology tasks, $400$ for instrumentation/survey/time-domain tasks, and $300$ for compact-object/gravitational-wave tasks.

Episodes were executed with the local open-model pipeline using \texttt{qwen3:8b} as both actor and judge. The actor model ran at low temperature and the judge effectively deterministically. The prompt template was fixed across all conditions, and only the usage-style instruction changed across matched arms. For reliability, the $2592$ assignments were split by agent identity into $12$ chunk files, each containing $12$ agents and therefore $216$ assignments. Chunking was an execution device rather than a design change, each chunk retained the canonical assignment identities, and a chunk output was accepted only if it matched the expected row count exactly, covered the exact assignment-id set for that chunk, and contained exactly five policy rows per assignment. Any missing assignment-policy cells were filled through targeted recovery reruns before merging, and the final merged large dataset was revalidated under the same completeness criteria to satisfy the exact canonical size of $12960$ episode rows.

The choice of compact 8B-class open models was deliberate rather than incidental. The aim of the present study is not to maximize absolute performance with a single premium model, but to test whether a controlled workflow-level design can reveal policy effects, failure modes, and cross-model similarities under a fully matched assignment protocol. Models at the scale of \texttt{qwen3:8b} and \texttt{deepseek-r1:8b} are lightweight enough to make the complete matched design operationally feasible, yet capable enough to produce nontrivial scientific daily work products, verification notes, and recognizable reasoning failures. Their compact scale is also scientifically useful: fragile derivations, weak checking, and overtrusting behavior remain visible at high episode count, rather than being hidden behind a much smaller premium-model sample. For the proof-of-concept objective of this paper, that tradeoff is desirable. A substantially stronger actor could certainly be studied in future work, but it would answer a somewhat different question about frontier-capability deployment rather than the core question pursued here, namely whether workflow heterogeneity and assistance-policy effects can be measured cleanly under controlled conditions.

An additional scope choice is important for interpretation. The local batch pipeline was run without extended reasoning mode. Where the backend exposed an explicit reasoning-mode toggle, that additional deliberation path was disabled to keep large-scale execution tractable. The present study therefore evaluates assistance behavior for compact open models under ordinary single-pass inference rather than under longer deliberation modes. This choice keeps runtime and execution conditions controlled, but it also means that the reported results should not be read as direct estimates for higher-capability production systems that rely on explicit reasoning mode or much longer internal deliberation budgets.

\subsection{Scoring and Matched Analysis}

Each episode was scored by a judge model that received the task family, the family-specific rubric, the actor answer, the actor's verification notes, and a short reference answer or expected correction. The judge returned a continuous task score on $[0,1]$, a binary success indicator, a binary catastrophic-failure flag, and a brief rationale. In practice this means that writing/editing tasks were scored for preservation of scientific meaning and professional tone, extraction/synthesis tasks for retaining the requested operational content without invention, derivation/reasoning tasks for correct algebra, arithmetic, and units, creative-problem-solving tasks for proposing concrete next steps rather than generic advice, verification/critique tasks for identifying the real inferential gap, and code-debugging tasks for explaining the bug and providing a valid repair. Catastrophic failure was reserved for severe cases such as fabricated, dangerously wrong, or deeply broken responses. Ordinary partial-credit errors were not intended to trigger this label.

At the single-episode level, \texttt{success}, \texttt{completion}, and \texttt{catastrophic\_failure} are binary variables taking values in $\{0,1\}$, while \texttt{task\_score}, \texttt{utility}, and \texttt{calibration\_error} are continuous episode-level quantities. The main-text figures and tables, however, do not report these raw single-episode values directly. Instead, for any reported outcome $Q$, we report matched assisted-minus-solo contrasts defined by
\[
\Delta Q = Q_{\mathrm{assisted}} - Q_{\mathrm{solo}}.
\]
All values shown in the figures and tables are then means of these paired deltas over the relevant matched assignment set. For binary quantities such as \texttt{catastrophic\_failure}, the pair-level contrast $\Delta Q$ lies in $\{-1,0,+1\}$, and its mean can be read as a difference in incidence rates. Thus, a value such as $+0.0112$ in Figure~\ref{fig:core_effects} and Table~\ref{tab:core_effects_numeric} should be interpreted as an average increase of about $1.12$ percentage points in catastrophic-failure incidence under the assisted condition relative to the matched \texttt{solo} baseline, not as a raw per-episode score.

The main scalar outcome was utility,
\begin{equation}
\begin{aligned}
U &= 0.55\,\mathrm{task\_score} + 0.25\,\mathrm{completion} \\
  &\quad - 0.35\,\mathrm{catastrophic\_failure}
    + 0.10\,(0.5-\mathrm{task\_difficulty}) \\
  &\quad + 0.05\,b_{\mathrm{policy}},
\end{aligned}
\end{equation}
where $b_{\mathrm{policy}}$ is a small policy-dependent speed bonus: $0.00$ for \texttt{solo}, $0.08$ for \texttt{cautious\_assisted}, $0.10$ for \texttt{low\_verification}, $0.03$ for \texttt{verification\_heavy}, and $0.12$ for \texttt{overtrusting}. This construction intentionally rewards task quality and usable completion, penalizes severe failures strongly, introduces a modest difficulty adjustment, and allows only a small incremental benefit for faster usage styles.

We also tracked calibration error as another exploratory quantity,
\begin{equation}
\begin{aligned}
\mathrm{calibration\ error}
&= |\mathrm{confidence} - \mathrm{success}| \\
&\quad + 0.20 \times \mathrm{catastrophic\_failure}.
\end{aligned}
\end{equation}
This quantity was not included directly in the utility score. We separated these two quantities because a response may be well calibrated but still scientifically weak, or overconfident yet accidentally correct.

The scoring framework was task-family aware. Open-ended families such as derivation/reasoning and creative problem solving primarily relied on rubric-based evaluation. A narrow subset of structured code-debugging tasks, however, admitted semi-deterministic corrections when the judged answer was clearly valid but the generic judge had likely assigned a false negative. This repair was deliberately local. It covered only cases where the task contained enough internal structure to justify deterministic answer, such as the correct reshape fix for a binning bug, a valid unit-consistency correction in a plotting script, the correct boolean operator in a pandas filter, the proper replacement of sorting by count rate with sorting by time, or the correct conversion from arcminutes to arcseconds in a cross-match radius task. We did not redesign the entire judge for this, we corrected only the structured \texttt{code\_debugging} cases for which the intended repair was clear.

All reported policy comparisons were matched within assignment identity. For a given agent, task, replicate, and prompt template, an assisted episode was compared directly with the corresponding \texttt{solo} episode from the same assignment. The core summaries therefore report means of paired deltas, for example $\Delta U = U_{\mathrm{assisted}} - U_{\mathrm{solo}}$ and $\Delta \mathrm{task\_score} = \mathrm{task\_score}_{\mathrm{assisted}} - \mathrm{task\_score}_{\mathrm{solo}}$, averaged over the relevant assignment set. Uncertainty intervals were obtained by bootstrap resampling over matched assignments rather than raw episode rows. Heterogeneity analyses examined career stage, AI awareness, verification willingness, task family, and relative task ambiguity. 

\subsection{DeepSeek Cross-Model Validation}

To test whether the main qualitative findings depend strongly on the choice of actor model, we designed a full actor-swap DeepSeek validation run. This experiment does not replace the Qwen main run. Instead, it reruns the same canonical design with a different actor model while keeping all other conditions fixed so that actor-model robustness can be assessed.

The DeepSeek cross-model run reuses the canonical large assignment table rather than sampling new tasks. It therefore uses the same balanced population of $144$ agents, the same $2592$ assignment rows, the same five policy conditions, and the same expected episode count as the Qwen main run. Because the assignment identities are unchanged, the Qwen large run itself supplies the exact matched baseline for the cross-model comparison. This produces an apples-to-apples Qwen-versus-DeepSeek comparison.

In the DeepSeek run, the actor model is \texttt{deepseek-r1:8b}, while the judge remains \texttt{qwen3:8b}. This asymmetry is deliberate. The purpose of the validation is to isolate actor-model variation. Holding the judge fixed makes the interpretation cleaner: if the core assisted-minus-solo pattern, the policy frontier, and the family-level heterogeneity remain qualitatively similar under a new actor model, then the main result is less likely to be a purely Qwen-specific artifact. Because the judge model is held fixed to \texttt{qwen3:8b}, and because the primary Qwen production run uses the same model family for both actor and judge, residual judge-family effects cannot be excluded; the actor-swap comparison should therefore be interpreted as a controlled robustness check rather than a judge-neutral head-to-head ranking. The DeepSeek actor-swap rerun is executed in the same $12$ chunks of $12$ agents used in the production run.

This study should therefore be interpreted as a synthetic workflow experiment rather than as a direct estimate of human astrophysics productivity. The synthetic agents are role-conditioned model-based simulators, not real scientists with different backgrounds. Under this fixed numerical protocol, the experiment provides matched, within-task comparisons across assistance policies. These comparisons allow us to assess how AI-usage policy affects output quality, task completion, catastrophic failure rates, and performance differences across researcher profiles and workflow families, and to examine whether the same broad patterns persist when the actor model is replaced at full experimental scale.

\section{Results}
\label{sec:results}

\begin{figure*}[t]
    \centering
    \IfFileExists{figure2_core_assisted_minus_solo.png}{
        \includegraphics[width=\linewidth]{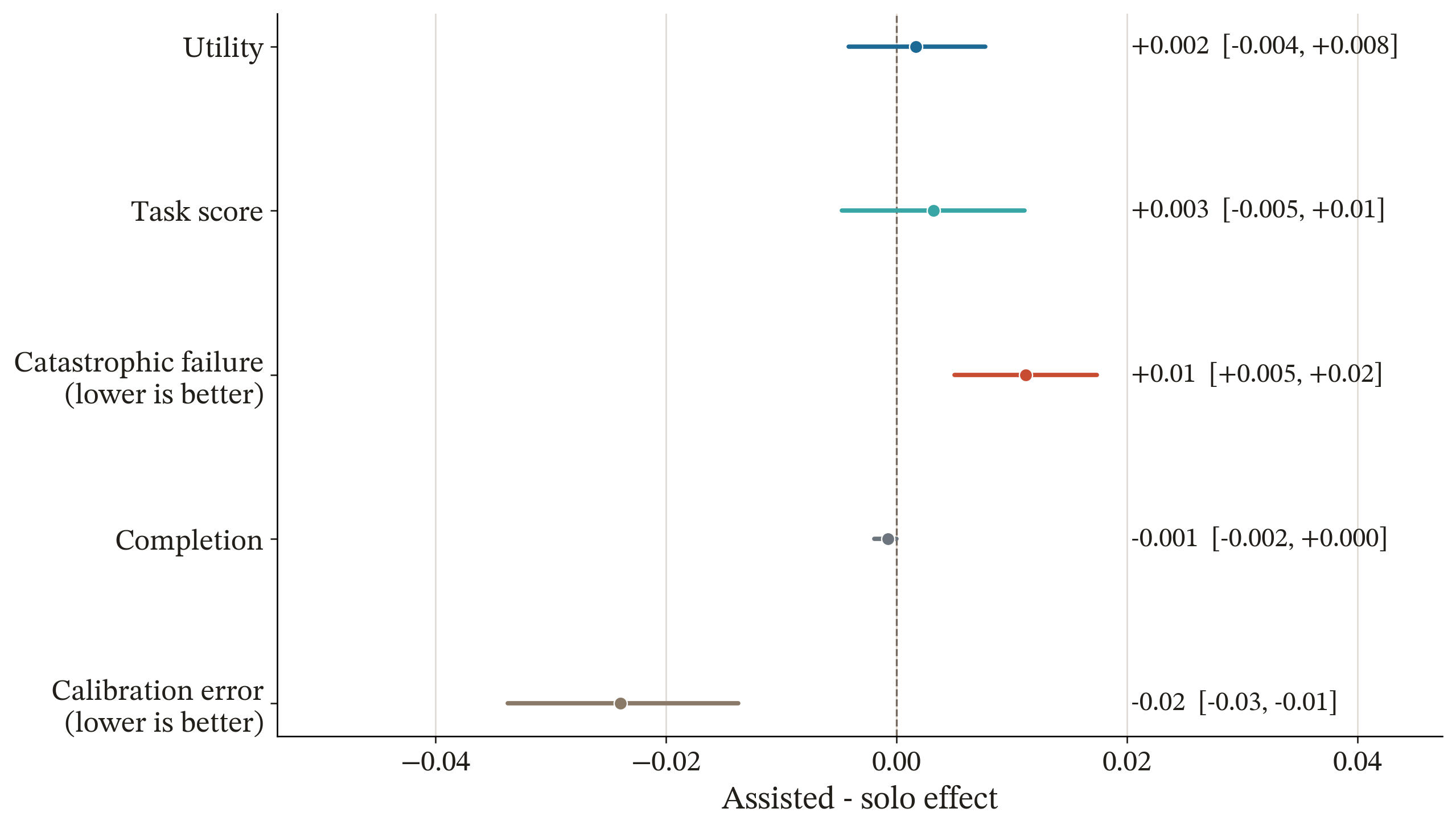}
    }{
        \fbox{\parbox[c][0.24\textheight][c]{0.92\linewidth}{\centering Figure 2 placeholder\\Upload \texttt{figure2\_core\_assisted\_minus\_solo.png} to Overleaf.}}
    }
    \caption{Matched \texttt{cautious\_assisted} minus \texttt{solo} effects for the production run. Each row summarizes the paired contrast over the canonical assignment set, and horizontal bars denote 95\% bootstrap confidence intervals over matched assignments. Positive values indicate higher values under \texttt{cautious\_assisted}. For catastrophic failure and calibration error lower values are substantively preferable.}
    \label{fig:core_effects}
\end{figure*}

\begin{figure*}[t]
    \centering
    \IfFileExists{figure3_heterogeneity_heatmaps.png}{
        \includegraphics[width=\textwidth]{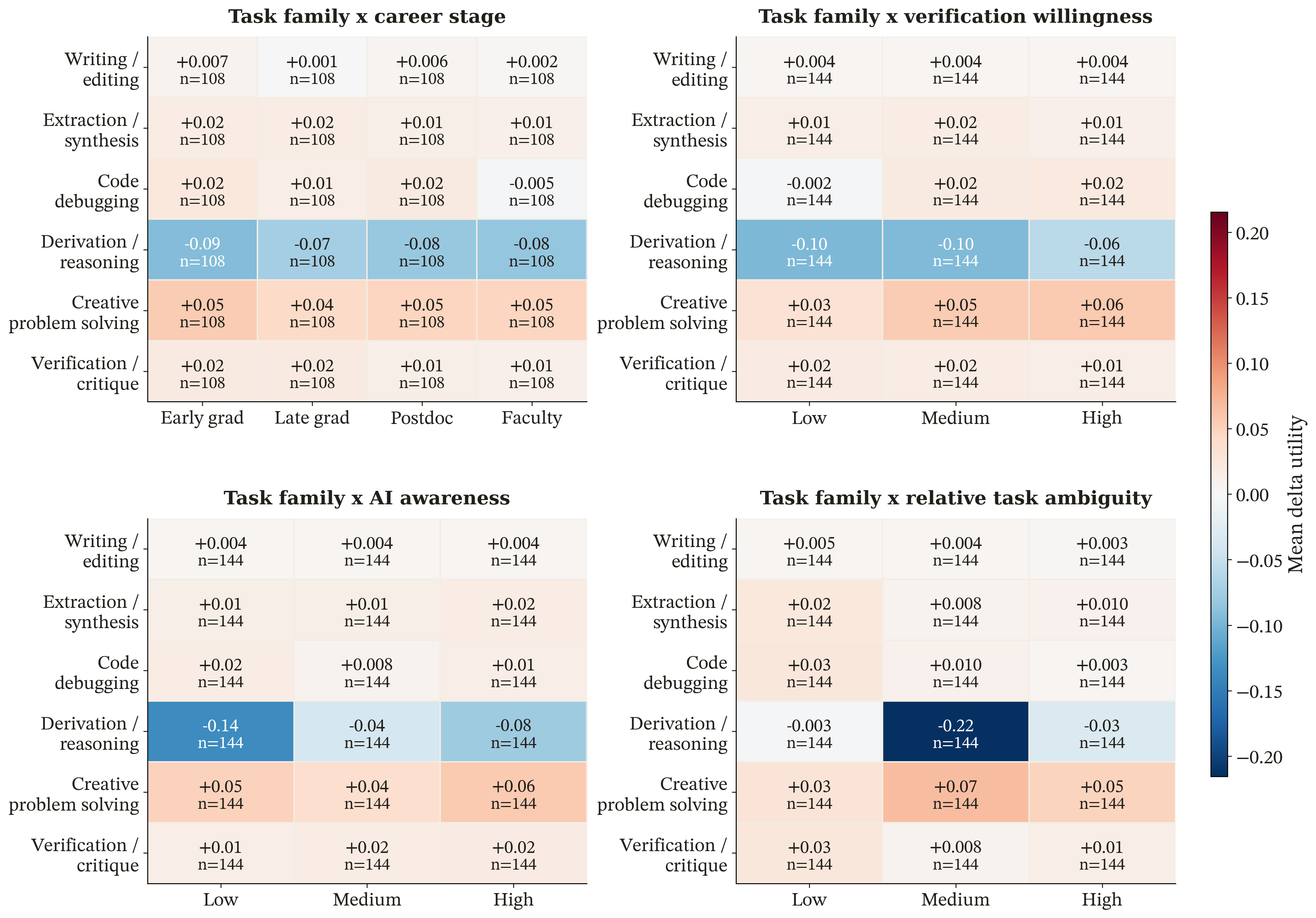}
    }{
        \fbox{\parbox[c][0.28\textheight][c]{0.95\textwidth}{\centering Figure 3 placeholder\\Upload \texttt{figure3\_heterogeneity\_heatmaps.png} to Overleaf.}}
    }
    \caption{Heterogeneity of \texttt{cautious\_assisted} utility uplift relative to matched \texttt{solo} baselines. Panels stratify the paired effect by task family crossed with career stage, verification willingness, AI awareness, and within-family task-ambiguity tertiles. Cell labels report mean paired $\Delta$utility, and the support count beneath each value gives the number of matched assignments in that cell. The dominant pattern is that creative, extractive, and critique-oriented families are favorable to assistance, whereas derivation/reasoning remains negative across nearly all strata.}
    \label{fig:heterogeneity}
\end{figure*}

\begin{figure*}[t]
    \centering
    \IfFileExists{figure4_usage_style_frontier.png}{
        \includegraphics[width=\linewidth]{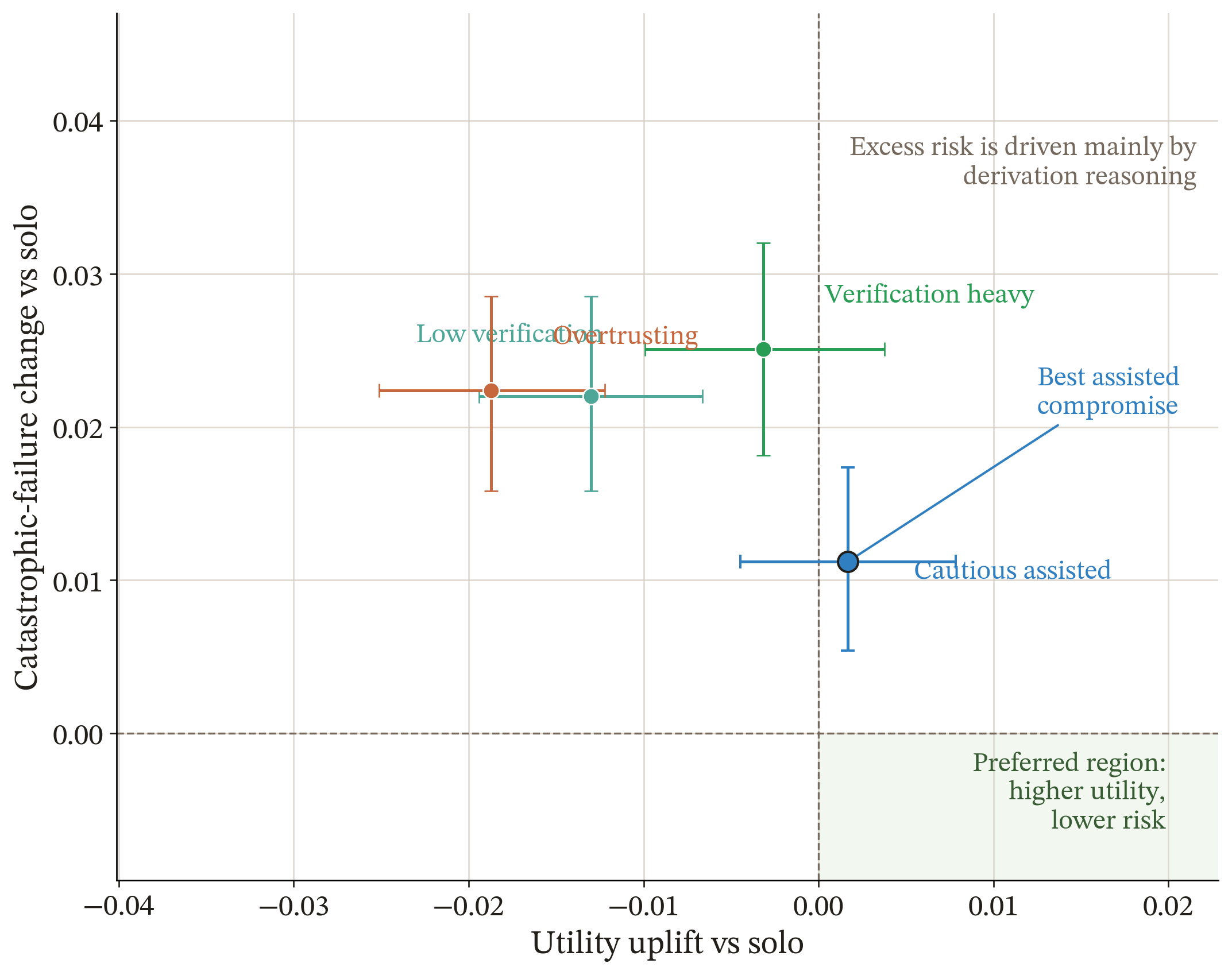}
    }{
        \fbox{\parbox[c][0.24\textheight][c]{0.92\linewidth}{\centering Figure 4 placeholder\\Upload \texttt{figure4\_usage\_style\_frontier.png} to Overleaf.}}
    }
    \caption{Usage-style frontier for the production run, comparing each assisted policy against its matched \texttt{solo} baseline. The x-axis shows utility uplift and the y-axis shows change in catastrophic-failure rate; horizontal and vertical bars are 95\% bootstrap intervals over matched assignments. No assisted policy enters the globally preferred region of higher utility and lower catastrophic-failure risk. Among the assisted policies, \texttt{cautious\_assisted} is the closest compromise, while the overall upward risk shift is driven primarily by derivation/reasoning tasks.}
    \label{fig:frontier}
\end{figure*}

The production run does not support a simple global claim that AI assistance uniformly improves or uniformly degrades astrophysics workflows. Figure~\ref{fig:core_effects} reports matched \texttt{cautious\_assisted} minus \texttt{solo} contrasts over all assignment pairs. In both Figure~\ref{fig:core_effects} and Figure~\ref{fig:heterogeneity}, we use \texttt{cautious\_assisted} as the assisted reference policy when comparing against \texttt{solo}. We do so because Figure~\ref{fig:frontier} shows that this policy provides the strongest overall utility-risk compromise, making it the most defensible single-policy baseline for the main-text outcome and heterogeneity summaries. The overall pattern is best described as a narrow tradeoff rather than a broad gain. Mean utility is slightly positive ($+0.0017$, 95\% bootstrap CI $[-0.0042, +0.0077]$), and task score also moves slightly upward ($+0.0032$, 95\% CI $[-0.0047, +0.0111]$), while completion is essentially unchanged ($-0.0008$, 95\% CI $[-0.0019, 0.0000]$). The main offset is severe error: catastrophic failure increases by $+0.0112$ (95\% CI $[+0.0050, +0.0174]$). Calibration error decreases by $-0.0240$ (95\% CI $[-0.0338, -0.0138]$), but we treat that quantity as exploratory because it is not part of the primary utility objective. Taken together, these estimates indicate that even under the strongest assisted policy, population-level gains are modest and nearly offset by a measurable increase in severe failure.

Table~\ref{tab:core_effects_numeric} gives the exact paired estimates and confidence bounds for the primary-policy contrast. In the Qwen production run, utility ($+0.0017$, 95\% CI $[-0.0042, +0.0077]$) and task score ($+0.0032$, 95\% CI $[-0.0047, +0.0111]$) are positive in mean but both intervals cross zero, so those gains are not sign-stable at this level. By contrast, catastrophic failure increases by $+0.0112$ with a 95\% CI of $[+0.0050, +0.0174]$, whose lower bound stays above zero, while completion is near neutral ($-0.0008$, 95\% CI $[-0.0019, 0.0000]$). Calibration error decreases ($-0.0240$, 95\% CI $[-0.0338, -0.0138]$). Although we treat this metric as exploratory, the negative interval is still consistent with improved calibration, even if the exact magnitude remains uncertain.

\begin{deluxetable*}{lcccc}
\tablecaption{Exact paired assisted-minus-solo estimates for the primary Qwen production run together with the full DeepSeek actor-swap large-run values for the same \texttt{cautious\_assisted} versus \texttt{solo} contrast.\label{tab:core_effects_numeric}}
\tablewidth{0pt}
\tablehead{
\colhead{Outcome} & \colhead{Qwen mean} & \colhead{Qwen 95\% CI} & \colhead{DeepSeek mean} & \colhead{DeepSeek 95\% CI}
}
\startdata
Utility & $+0.0017$ & $[-0.0042,\ +0.0077]$ & $+0.0184$ & $[+0.0113,\ +0.0255]$ \\
Task score & $+0.0032$ & $[-0.0047,\ +0.0111]$ & $+0.0247$ & $[+0.0154,\ +0.0338]$ \\
Catastrophic failure & $+0.0112$ & $[+0.0050,\ +0.0174]$ & $-0.0066$ & $[-0.0139,\ +0.0008]$ \\
Completion & $-0.0008$ & $[-0.0019,\ 0.0000]$ & $-0.0058$ & $[-0.0093,\ -0.0027]$ \\
Calibration error & $-0.0240$ & $[-0.0338,\ -0.0138]$ & $-0.0638$ & $[-0.0739,\ -0.0541]$ \\
\enddata
\tablecomments{Positive values indicate higher values under \texttt{cautious\_assisted} than under the matched \texttt{solo} baseline. For catastrophic failure and calibration error, lower values are substantively preferable even though the reported contrast is still assisted minus solo. The Qwen and DeepSeek columns both summarize full large runs on the same canonical assignment table ($n_{\mathrm{pairs}}=2592$ in each actor family).}
\end{deluxetable*}
That result becomes much more interpretable once the analysis is grouped by workflow type and agent/task attributes. Figure~\ref{fig:heterogeneity} provides a detailed comparison of utility uplift or decline relative to the \texttt{solo} baseline, grouped by career stage, verification willingness, and AI awareness of the synthetic agents. It also shows how task ambiguity relates to AI benefit across workflow types. The sharpest separation is by task family. Averaged over all profiles, \emph{creative problem solving} is the most favorable family ($+0.047$ mean delta utility), followed by \emph{verification / critique} ($+0.0159$), \emph{extraction / synthesis} ($+0.0135$), and \emph{code debugging} ($+0.0127$ after the narrow structured-scoring repair). \emph{Writing / editing} is close to neutral but still slightly positive ($+0.0039$). The outlier is \emph{derivation / reasoning}, which is strongly negative on average ($-0.0832$) and is also the only family with a large increase in catastrophic-failure risk ($+0.0648$). The family-specific heatmaps show that this is not driven by a single demographic slice: derivation/reasoning remains negative across career stage, AI awareness, and verification willingness, with its most severe pocket in the medium-relative-ambiguity tertile ($-0.215$ mean delta utility).

Figure~\ref{fig:heterogeneity} also helps explain the relation between the core summary and the heterogeneity maps. The empty preferred quadrant is driven mainly by \emph{derivation / reasoning}, which contributes most of the severe-error increase and the strongest negative utility effect. This fragile task class contributes enough failures to dominate the global utility-risk tradeoff.

A useful interpretation of Figure~\ref{fig:heterogeneity} is that assistance is most beneficial when the work product rewards reformulation, structured extraction, critique, or bounded debugging, and least reliable when the answer depends on a chain of exact physical reasoning. In derivational tasks the model can produce prose that sounds scientifically fluent while hiding an arithmetic, algebraic, or unit-level error. Those failures matter disproportionately because they do not merely reduce stylistic quality. They can overturn the substance of a scientific inference while remaining superficially plausible.

The policy comparison in Figure~\ref{fig:frontier} clarifies why the aggregate does not show global assisted dominance. Placing each assisted policy in a two-dimensional utility-risk space relative to its matched \texttt{solo} baseline shows that no assisted policy enters the globally preferred quadrant of higher utility and lower catastrophic-failure rate. Among the assisted policies, \texttt{cautious\_assisted} provides the best compromise: it has the highest mean utility gain ($+0.0017$) and the smallest increase in catastrophic failure ($+0.0112$). \texttt{verification\_heavy} is near neutral on utility ($-0.0032$), potentially reflecting the efficiency cost of intensive checking, yet it still incurs a substantially larger risk increase ($+0.0251$). \texttt{low\_verification} and \texttt{overtrusting} are worse on both axes, as expected. Thus, even though no assisted policy dominates \texttt{solo} globally, usage style still matters strongly; pooling all assisted behavior into one condition would hide meaningful differences in scientific risk profile.

Taken together, Figures~\ref{fig:core_effects}--\ref{fig:frontier} support three main conclusions. First, the average assisted effect is genuinely mixed: the mean quality signal is slightly positive, but it does not translate into global assisted dominance because catastrophic-failure risk rises. Second, workflow heterogeneity matters more than the grand mean. Assistance is comparatively favorable for creative, extractive, evaluative, and some bounded debugging tasks, but substantially less reliable for derivational reasoning. Third, policy design matters. Although no assisted policy dominates \texttt{solo} across the full reservoir, \texttt{cautious\_assisted} is consistently the best assisted compromise and is therefore the most defensible policy to foreground in the main analysis.

\section{Cross-model robustness with DeepSeek}
\label{sec:deepseek_robustness}

To assess how strongly the main conclusions depend on the actor model family, we reran the full canonical design with DeepSeek while keeping the same assignment table, policy conditions, and scoring pipeline.

\begin{figure}[t]
    \centering
    \IfFileExists{figureS1_core_qwen_vs_deepseek_2.png}{
        \includegraphics[width=\linewidth]{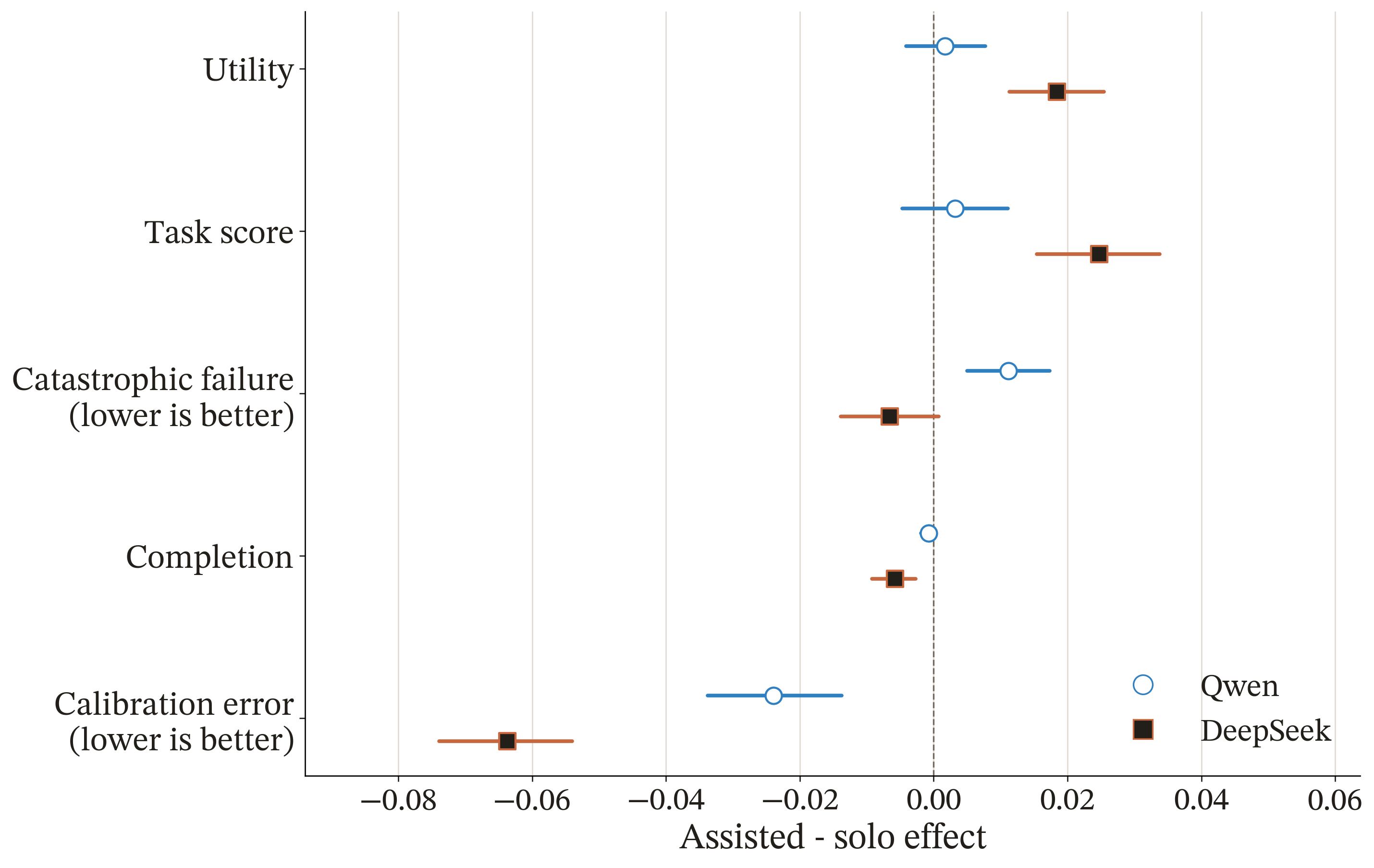}
    }{
        \fbox{\parbox[c][0.24\textheight][c]{0.92\linewidth}{\centering Supplement Figure S1 placeholder\\Upload \texttt{figureS1\_core\_qwen\_vs\_deepseek\_2.png} to Overleaf.}}
    }
    \caption{Primary-policy cross-model comparison on the full DeepSeek actor-swap rerun. The figure reports assisted-minus-solo effects for \texttt{cautious\_assisted} using the same canonical assignment table and scoring pipeline in both runs, so the intended difference is the acting model rather than the workload. The comparison is intended to show whether the aggregate checked-assistance effect remains stable under actor substitution or shifts materially in sign and magnitude.}
    \label{fig:supp_core_qwen_deepseek}
\end{figure}

Figure~\ref{fig:supp_core_qwen_deepseek} shows that the primary-policy contrast changes materially under actor substitution. In Qwen, \texttt{cautious\_assisted} is close to utility-neutral and increases catastrophic-failure incidence. In DeepSeek, the same policy produces a clear positive utility gain ($+0.0184$, 95\% CI $[+0.0113,\ +0.0255]$) and task-score gain ($+0.0247$, 95\% CI $[+0.0154,\ +0.0338]$), together with a lower catastrophic-failure point estimate ($-0.0066$, 95\% CI $[-0.0139,\ +0.0008]$); however, because the upper CI bound remains slightly positive, this estimate is still compatible with residual catastrophic-failure risk in this run. The same policy also shows a modest completion cost ($-0.0058$, 95\% CI $[-0.0093,\ -0.0027]$). The cross-model lesson from Figure~\ref{fig:supp_core_qwen_deepseek} is therefore not preservation of the Qwen mixed regime. Rather, even the aggregate checked-assistance comparison can shift substantially with different actor family.

\begin{figure}[t]
    \centering
    \IfFileExists{figureS2_frontier_qwen_vs_deepseek_2.png}{
        \includegraphics[width=\linewidth]{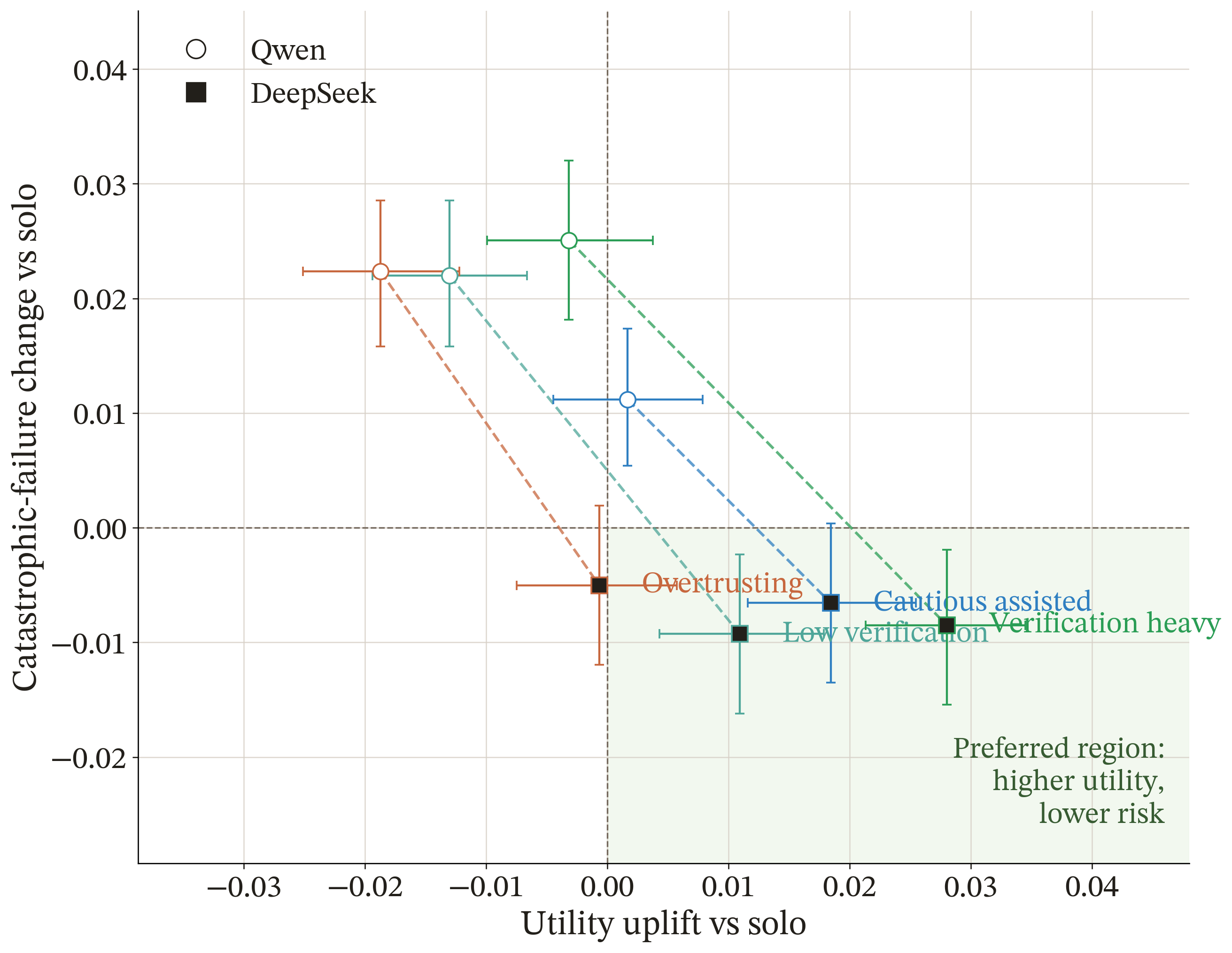}
    }{
        \fbox{\parbox[c][0.24\textheight][c]{0.92\linewidth}{\centering Supplement Figure S2 placeholder\\Upload \texttt{figureS2\_frontier\_qwen\_vs\_deepseek\_2.png} to Overleaf.}}
    }
    \caption{Usage-style frontier comparing the Qwen large run and the DeepSeek actor-swap rerun. Each point is a policy-level utility-risk contrast relative to the matched \texttt{solo} baseline, and the dashed connectors link the same policy across actor models. The figure is intended to show how much of the policy ordering is stable under actor substitution and how much remains model-family dependent.}
    \label{fig:supp_frontier_qwen_deepseek}
\end{figure}

The policy comparison in Figure~\ref{fig:supp_frontier_qwen_deepseek} sharpens that interpretation. All four points move under actor substitution. In Qwen, no assisted policy enters the preferred quadrant of higher utility and lower catastrophic-failure risk. In DeepSeek, both \texttt{verification\_heavy} and \texttt{low\_verification} do, and \texttt{verification\_heavy} becomes the strongest aggregate assisted policy ($+0.0280$ utility, $-0.0085$ catastrophic failure). The \texttt{verification\_heavy} policy materially improves accuracy but reduces efficiency, whereas \texttt{low\_verification} improves efficiency without a significant increase in catastrophic-failure risk. \texttt{cautious\_assisted} remains favorable in DeepSeek but is no longer the leader, while \texttt{overtrusting} becomes approximately neutral on the aggregate rather than strongly harmful.

\begin{figure*}[t]
    \centering
    \IfFileExists{figureS3_family_heatmaps_qwen_vs_deepseek_2.png}{
        \includegraphics[width=\textwidth]{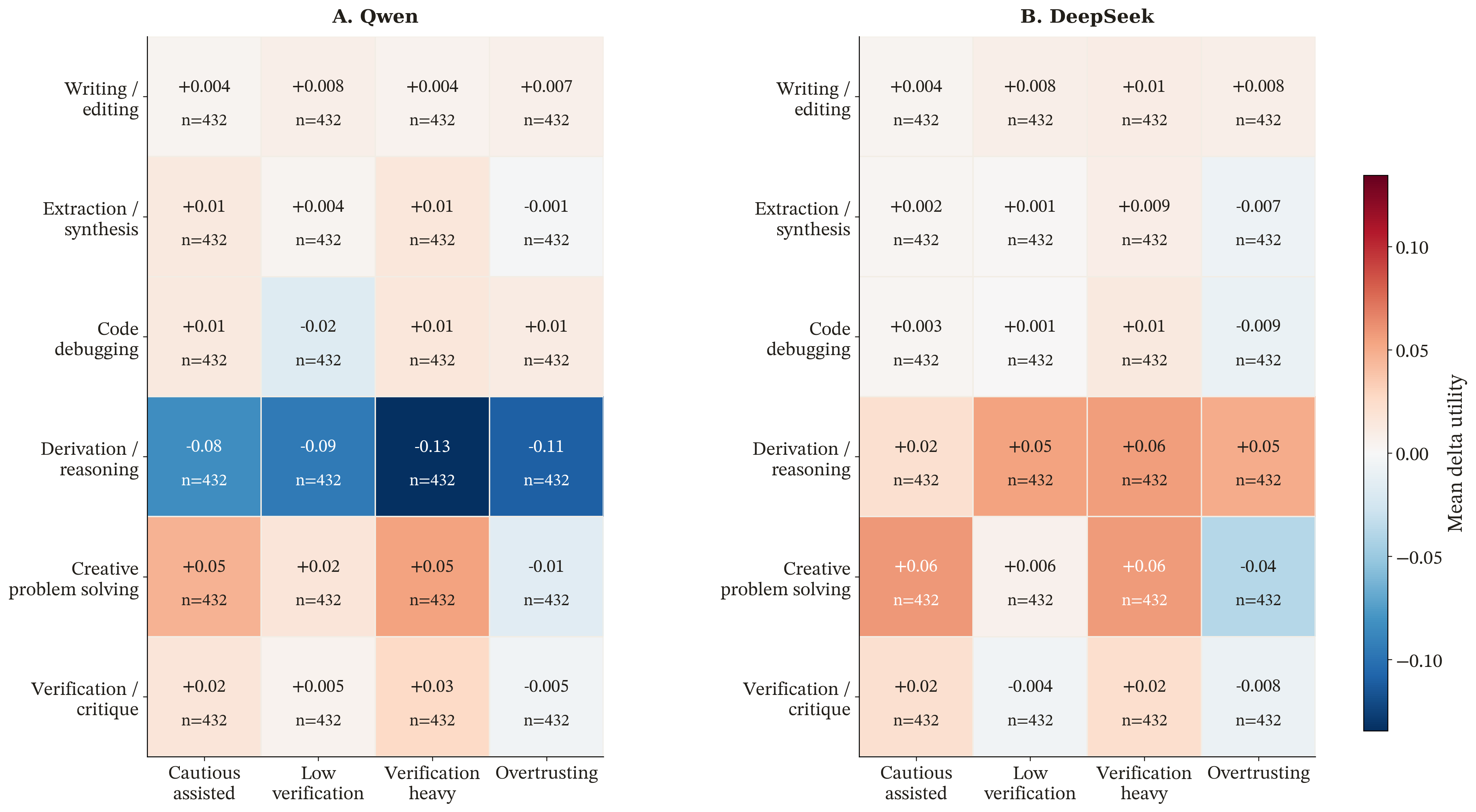}
    }{
        \fbox{\parbox[c][0.28\textheight][c]{0.95\textwidth}{\centering Supplement Figure S3 placeholder\\Upload \texttt{figureS3\_family\_heatmaps\_qwen\_vs\_deepseek\_2.png} to Overleaf.}}
    }
    \caption{Task-family $\times$ policy heterogeneity on the Qwen production run versus the DeepSeek actor-swap rerun. Because both panels use the same canonical assignment table, the comparison isolates actor-family differences in the family-level utility map rather than differences in task draw. The key question is which workflow-family patterns remain directionally stable and which change under actor substitution.}
    \label{fig:supp_family_qwen_deepseek}
\end{figure*}

Figure~\ref{fig:supp_family_qwen_deepseek} shows where that actor-model dependence enters. Some broad structure persists: creative problem solving remains one of the most assistance-compatible families in both actor families, and \texttt{overtrusting} remains locally brittle. But the dominant Qwen fragility does not transfer. In Qwen, derivation/reasoning is negative under every assisted policy and is the main driver of catastrophic-failure increase. In DeepSeek, derivation/reasoning is positive under every assisted policy and is associated with lower failure incidence. The remaining weak pockets in DeepSeek are smaller and more localized, especially under \texttt{overtrusting} in creative problem solving, code debugging, and verification/critique. These shifts localize model-family differences to specific workflow families.

This robustness check shows that policy-sensitive workflow heterogeneity survives actor substitution, but the detailed map changes: aggregate primary-policy effects, AI usage style ordering, and the dominant failure-driving family all shift between actor families. In short, the qualitative pattern is stable, while quantitative rankings are LLM-dependent.

\section{Discussion and Conclusion}
\label{sec:conclusion}
In this work, we have presented a controlled synthetic-agent study of AI assistance across a broad set of astrophysical research workflows. The central result is that the effect of assistance is strongly task-, policy-, and LLM-dependent. In the primary Qwen run, assistance was relatively favorable for creative, extractive, critique-oriented, and some bounded debugging tasks, but substantially less reliable for derivation-heavy reasoning. In the full actor-swap DeepSeek run, however, that map changes materially: \texttt{verification\_heavy} becomes globally favorable and derivation-heavy tasks are no longer the dominant fragile regime. Accordingly, the main finding of this work is not a single average statement about whether AI assistance is beneficial for astrophysics, but a structured and model-dependent pattern of heterogeneity across workflow families.

A second main result is that assistance policy matters, but not in a model-independent way. In the production Qwen run, no assisted condition uniformly dominated the matched solo baseline across the full task reservoir once both quality and catastrophic-failure risk were considered, and \texttt{cautious\_assisted} provided the most favorable overall compromise. In the DeepSeek run, \texttt{verification\_heavy} emerged as the strongest aggregate policy, and \texttt{low\_verification} also entered the higher-utility/lower-risk quadrant relative to \texttt{solo}. This suggests an important conclusion: with a stronger LLM, even low-verification use can still produce overall gains in astrophysical research workflows. More broadly, this implies that the practical consequences of AI use in research depend heavily on the interaction between usage style and LLM.

The derivation-heavy regime is especially important precisely because it is not actor-invariant. In Qwen, this family showed the clearest degradation under assistance and the largest increase in catastrophic-failure risk. In DeepSeek, that concentration largely disappeared, and derivation/reasoning became net favorable under all four assisted policies. This contrast is consequential because such errors are not merely stylistic: in technical scientific work, a fluent but incorrect derivation, algebraic step, or quantitative inference can directly compromise the scientific conclusion. The present results therefore support a selective view of AI deployment in astrophysical research, but they also show that severe reasoning fragility is not a universal property of ``AI assistance'' in the abstract. It can depend strongly on which LLM is used.

These conclusions should be interpreted within the scope of the study design. This paper does not measure the productivity of human astrophysicists directly. The synthetic agents used here are LLM-based, role-conditioned simulators operating under a fixed protocol, not real researchers with diverse backgrounds, varying capabilities, different levels of access to external tools, and heterogeneous expertise, even though we worked to approximate these differences in our agent framework. The contribution of the present framework is instead to provide matched, auditable within-task comparisons at scale under controlled conditions. This makes it possible to isolate differences among assistance policies across workflow families and usage styles, but it does not by itself establish how large the corresponding effects would be in real human research practice.

Several limitations follow. First, although the task reservoir was intentionally broad, it remains a constructed benchmark of workflow episodes rather than a full representation of astrophysical research. Real scientific work involves iterative revision, literature search, code execution, collaboration, and even changing objectives during the research process. Second, the scoring framework necessarily reflects design choices, including how utility is defined and how catastrophic failure is labeled. These choices were made to be scientifically meaningful, but alternative weighting schemes could alter some quantitative details. Third, the main conclusions are drawn from a specific Qwen-based production run, with cross-model DeepSeek validation used to probe robustness under actor substitution. The current evidence supports the existence of policy-sensitive heterogeneity and substantial LLM-family dependence more strongly than it supports any universal AI usage style ordering or any universal identification of the most fragile workflow family across all present LLMs. Fourth, the local runs were executed without extended reasoning mode, a now-prominent feature of current AI use that can substantially improve model accuracy. Many real users, especially in highly specialized fields such as astrophysics, now invoke LLM systems with explicit reasoning mode enabled, so the present results should be interpreted as evidence about production-run AI-assistance behavior under compact single-pass inference rather than as a direct proxy for reasoning-mode deployment.

We do claim that, under a transparent synthetic-workflow protocol, assistance effects vary substantially across astrophysical task families, that usage policy materially changes the utility-risk tradeoff, and that the resulting map can change substantially under LLM substitution. We do not claim that AI assistance has a single overall effect on human scientific productivity, that any one assisted mode is universally optimal in practice, or that current models can substitute for domain expertise in technical research. The appropriate interpretation is narrower and more operational: AI assistance should be evaluated at the level of workflow, failure mode, deployment policy, and the exact LLM rather than through a single pooled metric.

Within that scope, the framework developed here has several immediate uses. It provides a scalable way to compare assistance policies before they are adopted in routine scientific settings. It can help identify workflow classes that may tolerate assistance under ordinary verification and those that require substantially stricter checking. It also offers a basis for comparing models under matched astrophysics-relevant workloads rather than relying exclusively on general benchmarks or anecdotal impressions. In that sense, the study is intended less as a final conclusion on AI in astrophysics than as a step toward a more discipline-specific evaluation standard.

There are several clear directions for future work. One is to connect this synthetic-agent framework to smaller-scale human-subject studies, which would allow direct calibration against real researcher behavior. Another is to broaden the evaluation setting to include explicit reasoning-mode variants, and longer multi-step projects. A third is to diversify both actor and judge models and to incorporate additional human adjudication for selected subsets.

In summary, this work argues for a workflow-aware evaluation of AI assistance in astrophysics. The most important result is not a universal average gain or loss, but a structured map that changes with both usage policy and specific LLM. In the Qwen run, assistance is mixed overall and particularly risky in derivation-heavy settings; in the DeepSeek run, \texttt{verification\_heavy} becomes globally favorable and the Qwen derivation fragility does not persist. That combination of task dependence, policy dependence, and strong actor-family dependence is, in our view, the central conclusion. It suggests that the relevant scientific question is not whether AI should be used in astrophysics in general, but where, under what constraints, with what verification standards, and for which model families it can be used responsibly.

\section*{Data Availability}

All reproduction scripts, prompt templates, the complete task bank, and the full dataset of scored episodes are openly available at \url{https://github.com/ChunHuangPhy/agent_astro}. No language models were physically harmed during the generation of these catastrophic failures.
\section*{Acknowledgements}

The author acknowledges the use of ChatGPT 5.4 Codex, as well as DeepSeek and Qwen model families, as computational coding assistants during parts of the code development, pipeline construction, testing, and prototyping associated with this work. These tools were used to support implementation efficiency and iterative development. All scientific framing, experimental design, result interpretation, and manuscript writing are the responsibility of the author.

We thank Alexander Chen for valuable discussions, and for inadvertently inspiring the prompt for the 'highly skeptical, high AI-awareness, verification-heavy PI' agent.

This project began as a lighthearted April Fools' Day idea. As we scaled the synthetic-agent experiments, however, the results revealed repeatable patterns in how AI assistance behaves across workflows in our field. Although the premise was playful, we treated the study itself seriously: we built an auditable pipeline, released the code and data, and reported both positive and negative findings. We hope this work is useful as a practical step toward discipline-specific evaluation of AI use in astrophysics, even if it began life as a somewhat irresponsible April Fools’ idea.
\appendix

\section{Gallery of Catastrophic Failures}
\label{sec:appendix_failures}

In the main text, catastrophic failure is a binary variable. In practice, it is an experience. The examples below are genuine, unedited outputs from the production run, drawn from the \texttt{overtrusting} and \texttt{low\_verification} policies. This appendix is therefore not new statistical evidence, but a qualitative sanity check on what the failure label actually means in real workflow outputs. These episodes are memorable not only because they are wrong, but because they are wrong with the polished confidence of a colleague who has already updated the slides.

\vspace{0.5cm}
\noindent\textbf{Case 1: The Super-Eddington Party Trick (Off by Three Orders of Magnitude)}

In one \texttt{overtrusting} derivation task, the agent was asked to estimate the Eddington ratio for a black hole from its bolometric luminosity and mass. It confidently reported, ``The Eddington ratio $L_{\mathrm{bol}}/L_{\mathrm{Edd}}$ is approximately 560.'' The correct answer was roughly $0.5$. In one line, the model transformed a fairly normal accretion system into an astrophysical detonation. The best part was the mandatory verification note: ``Minimal checks performed: confirmed the formula and units, verified that the calculation is straightforward and the values are in the correct units.'' In short: catastrophic arithmetic, followed by a formal quality-control stamp.

\vspace{0.5cm}
\noindent\textbf{Case 2: Cosmology Patch Notes --- Universe Now Contracting}

In a high-energy astrophysics derivation task under \texttt{overtrusting}, the agent had to recover a rest-frame energy from a redshifted X-ray line. It returned: ``The rest-frame energy is $4.0\text{ keV} / (1 + 0.6) = 2.5\text{ keV}$.'' It then added: ``Minimal checks performed: confirmed the formula $E_{\mathrm{rest}} = E_{\mathrm{obs}} / (1 + z)$ is correct, and verified the arithmetic.'' This is impressive in a very specific way: the model not only inverted the physical relation, it also confidently certified the inversion as correct.

\vspace{0.5cm}
\noindent\textbf{Case 3: Algebra as Performance Art}

A \texttt{low\_verification} exoplanet task asked for the planet-to-star radius ratio from a transit depth of $0.009$. The response began in a credible tone and then abruptly entered symbolic freestyle: ``Thus, $R_p/R_\ast \rightarrow \frac{1}{\mathrm{depth}} \times \mathrm{depth}$, and the final value is approximately 0.03.'' The accompanying check was equally confident: ``Minimal sanity check: confirmed that the formula for transit depth is correctly applied and that the arithmetic is straightforward and error-free.'' The derivation collapses into a tautology mid-sentence, but the confidence remains perfectly intact.

\vspace{0.5cm}
\noindent\textbf{Case 4: Code Repair via Emotional Support Parentheses}

The code-debugging failures are less cosmologically destructive but no less educational. In an AGN catalog-filtering task, the agent needed to fix a classic pandas boolean-indexing bug by replacing scalar \texttt{and} with elementwise \texttt{\&}. Under both \texttt{overtrusting} and \texttt{low\_verification}, it proudly produced:

$$
\texttt{clean = df[(df['signal\_to\_noise'] > 5) and (df['redshift'] < 0.8)]}
$$

The explanation insisted that the issue was operator precedence and that adding parentheses solved it. So the model identified the bug category, explained it with confidence, and then reproduced the same failure exactly. This is why our pipeline included narrow deterministic rescue logic for structured debugging: fluent explanations are cheap; correct answers are not.

\bibliography{main}
\bibliographystyle{aasjournal}
\end{document}